\documentclass[11pt]{article}
\usepackage{moriond}

\def\be{\begin{equation}}
\def\ee{\end{equation}}
\def\bea{\begin{eqnarray}}
\def\eea{\end{eqnarray}}

\newcommand{\Photo}{}

\begin{document}
\vspace*{4cm}
\title{HIGHER ORDER PROTON LIFETIME ESTIMATES IN GRAND UNIFIED THEORIES}

\author{ H. KOLE\v{S}OV\'{A}}

\address{Faculty of Nuclear Sciences and Physical Engineering, Czech Technical University in Prague, B\v{r}ehov\'{a} 7,\\
Prague 114 19, Czech Republic}

\maketitle\abstracts{
Since the main experimentally testable prediction of grand unified theories is the instability of the proton, precise determination of the proton lifetime for each particular model is desirable. Unfortunately, the corresponding computation usually involves theoretical uncertainties coming e.g. from ignorance of the mass spectrum or from the Planck-suppressed higher-dimensional operators, which may result in errors in the proton lifetime estimates stretching up to several orders of magnitude. On the other hand, we present a model based on SO(10) gauge group which is subsequently broken by a scalar adjoint representation, where the leading Planck-suppressed operator is absent, hence the two-loop precision may be achieved.}

\section{Introduction}
Baryon number violation as a smoking gun of the grand unified theories (GUTs) is searched for by number of experiments. Let us concentrate for simplicity on the $p\to e^+\pi^0$ decay which has a clean signature in the water Cherenkov detectors and which is usually predicted by non-SUSY GUT models. For this channel the recent bound on the proton lifetime is set by the Super-Kamiokande experiment\cite{mine} to $\tau_p\geq 1.4\times 10^{34}\,$y, whereas the future Hyper-Kamiokande experiment assumes to reach the bound of $\tau_p\geq 10^{35}\,$y after 10 years of data taking\cite{mine}. If this progress on experimental side should distinguish between different GUT models, the uncertainties in the theorists' computations of proton lifetime can not exceed one order of magnitude and a next-to-leading order computation is necessary.

The aim of this text is twofold. First, we would like to explain in Section \ref{SecErrors} what are the ingredients for such a higher order computation of the proton lifetime.
Second, we would like to present a particular GUT model based on SO(10) gauge group which was analyzed at next-to-leading order and interesting correlations between the proton lifetime and particle spectrum were found (see Section \ref{SecSO10}).

\section{Error estimates for proton lifetime computation in GUTs}\label{SecErrors}
%
In non-SUSY GUTs the proton decay is predominantly mediated by vector bosons with large masses $M_X$ and if the the contributions of the scalar fields are neglected, $M_X$ is also the scale above that the $\beta$-functions for the SM couplings coincide. For this reason, we assume in the first approximation that the scale $M_G$ where the SM couplings intersect is equal to $M_X$, hence determines the proton lifetime $\tau_p$. The errors in the position of $M_G$ then lead to errors in $\tau_p$, let us examine what are their sources. 

\begin{enumerate}
\item\emph{Finite order of perturbation theory.}\label{2loop} Defining the variable $t=\frac{1}{2\pi}\log{\frac{\mu}{M_Z}}$, the renormalization group equation for the gauge couplings may be rewritten in the form
  \begin{equation}\label{RGE}
  \frac{\mathrm{d}}{\mathrm{d}t}\alpha^{-1}_i=-a_i-\frac{b_{ij}}{4\pi}\alpha_j - \frac{c_{ijk}}{16\pi^2}\alpha_j\alpha_k +\dots
  \end{equation}
where $\alpha_i\equiv g_i^2/4\pi$, $a$, $b$ and $c$ are $\mathcal{O}(1)$ coefficients corresponding to one-, two- and three-loop contributions, and the dots correspond to higher order contributions. Consequently, one can estimate that if working at one-loop level, the error coming from neglecting the two-loop effects at the GUT scale $M_G\sim 10^{16}\,\mathrm{GeV}$ is of the order of
\begin{equation}\label{error2loop}
\Delta \alpha_i^{-1}(M_G)_{\mathrm{1-loop}} \sim \frac{\mathcal{O}(1)}{16\pi^2} (t_G-t_Z)\sim 0.03 \times \mathcal{O}(1).
\end{equation}
Similarly, the error of the two-loop calculation reads
\begin{equation}\label{error3loop}
\Delta \alpha_i^{-1}(M_G)_{\mathrm{2-loop}} \sim \frac{\mathcal{O}(1)}{(16\pi^2)^2} (t_G-t_Z)\sim 0.0002 \times \mathcal{O}(1)\,.
\end{equation}
\item\emph{Measurement of $\alpha_{S}$ at the EW scale.}\label{alphS} Taking into account the experimental value for the strong coupling $\alpha_S(M_Z)=0.1185\pm0.0006$, $1\sigma$ uncertainty in $\alpha_S^{-1}$ reads
\begin{equation}\label{errorStrong}
\Delta \alpha_S^{-1}(M_G)_{\mathrm{1\sigma}} \approx \Delta \alpha_S^{-1}(M_Z)_{\mathrm{1\sigma}} = 0.04\,
\end{equation}
(strictly speaking, the uncertainty is reproduced at the GUT scale without any change only in case of one-loop running which is linear). Typically, the $\mathcal{O}(1)$ factor in Eq.~\ref{error2loop} makes the one-loop error more significant than the error in Eq.~\ref{errorStrong}, however, it is obvious that switching to the three-loop calculation is meaningless with today's precision in $\alpha_S$ measurement.
\item\emph{Threshold effects.}\label{ParMatching} 
In real models the masses of the heavy fields are not equal and so called matching\cite{Weinberg:1980wa} of the couplings at the GUT scale has to be performed. If a simple gauge group $G$ is spontaneously broken into a direct product of subgroups $G_i$ (with at most one abelian factor) at the scale $\mu$, then at the one-loop formula reads\cite{Bertolini:2013vta}
\begin{eqnarray}\label{matching}
\alpha_i^{-1}(\mu)&=&\alpha_G^{-1}(\mu) - 4\pi \lambda_i(\mu)\\
\nonumber\lambda_i(\mu) &=& \frac{S_2(V_i)}{48\pi^2}+\frac{1}{8\pi^2}\Biggl[-\frac{11}{3}S_2(V_i)\log\frac{M_{V}}{\mu} +\frac{4}{3}\kappa_F S_2(F_i) \log\frac{M_{F}}{\mu}+\frac{1}{3}\eta_S S_2(S_i) \log\frac{M_{S}}{\mu}\Biggr].
\end{eqnarray}
($V$, $F$ and $S$ denote the heavy vector bosons, fermions and scalars integrated out at the scale $\mu$).
As a rule, all the heavy masses are products of the scalar couplings and the vacuum expectation value of the scalar field responsible for the breaking of the group $G$, and the matching scale $\mu$ is chosen close to the barycenter of these masses. Consequently, the heavy masses $M$ should lie close to $\mu$ and comparing Eqs.~\ref{matching}~and~\ref{error2loop} one obtains
$$\Delta \alpha_i^{-1}(M_G)_{\mathrm{matching}} \sim \Delta \alpha_i^{-1}(M_G)_{\mathrm{1-loop}}.$$
if $\left|\log\frac{M}{\mu}\right| \sim \frac{1}{16\pi^2}\log\frac{M_G}{M_Z}$. However, the threshold effects may be even more significant for larger splitting of the heavy spectrum. Therefore, if one would like to work at two-loop precision level, the (one-loop) threshold effects have to be taken into account.
\item\emph{Gravity induced operators.}\label{ParGravity} Since the unification scale is typically rather close to the (reduced) Planck scale $M_{\mathrm{Pl}}=2.43\times 10^{18}$, one has to take into account also the effective operators such as\cite{Calmet:2008df}
    \begin{equation}\label{planck}
    \frac{k}{M_{\mathrm{Pl}}}\mathrm{Tr}\left(G_{\mu\nu}G^{\mu\nu}H\right)
    \end{equation}
    where $G_{\mu\nu}$ is the GUT field strength, $H$ is a scalar multiplet and $k$ is an $\mathcal{O}(1)$ constant. Let us suppose that the vacuum expectation value (VEV) $\langle H\rangle\equiv V_G$ breaks the gauge group $G$ to the SM. Then $V_G\sim M_G$ and Eq.~\ref{planck} contains the contributions to the SM kinetic terms $\epsilon_i \mathrm{Tr}\left(F_{\mu\nu}^iF^{i\mu\nu}\right)$ with $\epsilon_i\sim\mathcal{O}(10^{-2})$. In order to maintain the canonical normalization of the kinetic terms, one has to redefine $A_\mu^i\to(1+\epsilon_i)^{1/2}A_\mu^i$, $g_i\to(1+\epsilon_i)^{-1/2} g_i$. It is this redefined coupling which is measured at the EW scale, however, the unification condition holds for the original couplings: $\alpha_G=(1+\epsilon_i)\alpha_i(M_G)$ for all $i=1,2,3$. Naturally, this induces the uncertainty
    $$\Delta \alpha_i^{-1}(M_G)_{\mathrm{Pl}} \sim \mathcal{O}(10^{-2})$$
    which is comparable to the one-loop error, let alone the fact that in case of large number of fields in the theory, the effective Planck scale may be lowered by as much as one order of magnitude\cite{Calmet:2008df} and then the ``gravity smearing'' effects are even more pronounced.
\end{enumerate}

\begin{figure}
\begin{center}
\includegraphics[width= 0.8 \textwidth]{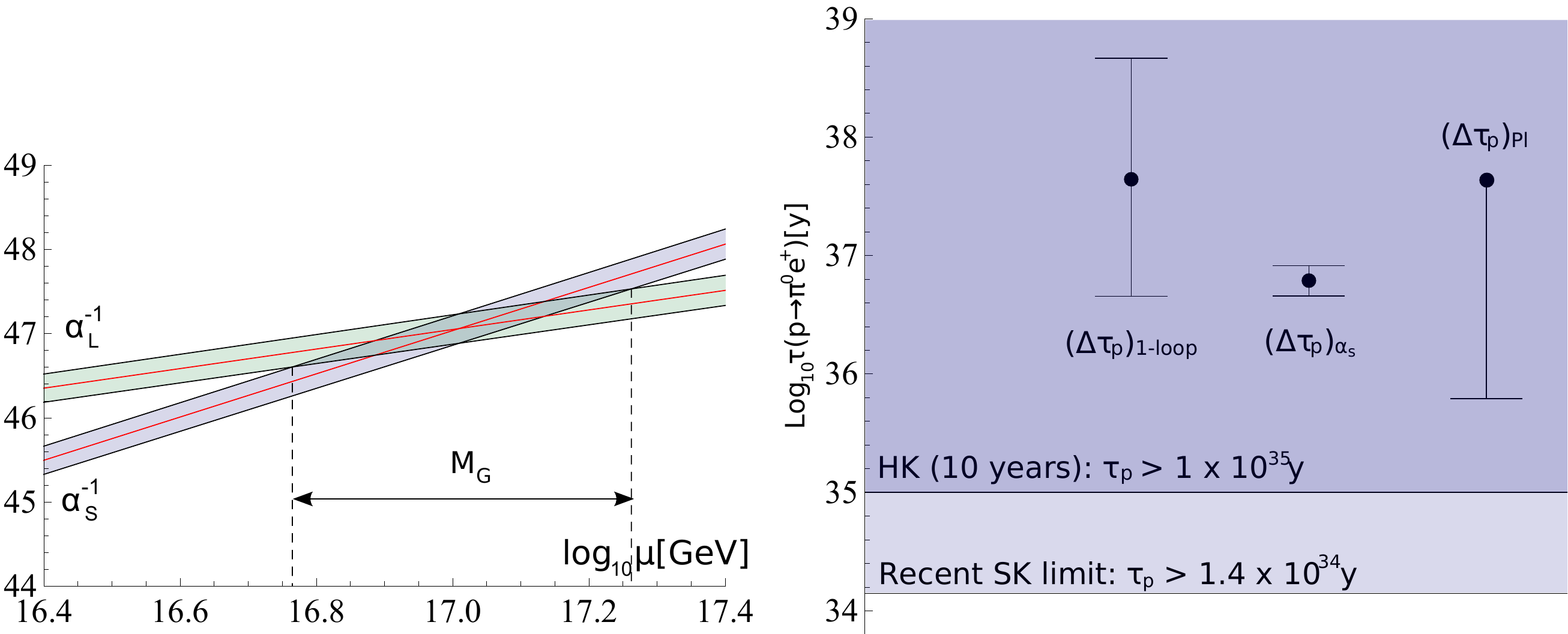}
\end{center}
\caption{On the l.h.s., the effect of $\Delta \alpha_i^{-1}(M_G)_{\mathrm{1-loop}}$ on the determination of $M_G$ is shown whereas the errors in the proton lifetime due to $\Delta \alpha_i^{-1}(M_G)_{\mathrm{1-loop}}$, $\Delta \alpha_S^{-1}(M_G)_{\mathrm{1\sigma}}$, and $\Delta \alpha_i^{-1}(M_G)_{\mathrm{Pl}}$ are depicted on the r.h.s.}
\label{FigProton}
\end{figure}

To demonstrate the effect of the uncertainties in $\alpha_i^{-1}$ on the proton lifetime prediction, let us consider a toy model where only the SM fields are present up to the unification scale $M_G$ which is determined by the intersection of the $SU(2)_L$ and strong couplings (this could be the case e.g. in the simplest flipped SU(5) models). Assuming $M_X=M_G$ let us evaluate the proton lifetime using the naive formula\footnote{We are aware of the fact that in the precise calculation the proper formula\cite{proton} has to be used, however, we believe that this simplification is enough for estimating the errors which are mostly driven by the fact that $\tau_p\propto M_G^4$}
$\tau_p \approx \frac{M_G^4}{g^4 m_p^5}.$
If the uncertainties $\Delta \alpha_i(M_G)^{-1}$ are applied, the corresponding errors in $M_G$ are found to be rather large due to the small angle between the two curves describing the running. On the l.h.s of Figure \ref{FigProton} we show an example of the effect of
$\Delta \alpha_i^{-1}(M_G)_{\mathrm{1-loop}}$ on the determination of $M_G$ (assuming the $\mathcal{O}(1)$ factor in Eq.~\ref{error2loop} to acquire values between $-5$ and $5$), the corresponding errors in the proton lifetime are then shown in the first column of the plot on the r.h.s. The second column in the same plot corresponds to the proton lifetime computed using the two-loop $\beta$-function with the error-bars depicting the 1$\sigma$ uncertainty in $\alpha_S(M_Z)$. Finally, the third column shows the error coming from the Planck-suppressed operators where we assumed $k\in(0,0.5)$, $\epsilon_L>0$ and $\epsilon_S<0$ which leads to a shift to lower $M_G$ and hence only one-sided error bar in $\tau_p$. For $M_G$ shifted to higher values, the gravity smearing effects get completely out of control and the two couplings may not even intersect.

Finally, let us add that the 30-40\% error in the lattice computation of the hadronic matrix elements\cite{Aoki:2013yxa} introduces also an uncertainty in $\tau_p$, which, however, does not exceed one order of magnitude. On the other hand, the ignorance of the superpartner spectrum in case of the SUSY GUTs may lead to an error stretching up to several orders of magnitude.

\section{Minimal SO(10)}\label{SecSO10}
Fortunately, the gravity smearing effects are absent in some of the unification models such as the one based on SO(10) gauge group broken by the VEV of a scalar adjoint representation $45_H$ since $\mathrm{Tr} \left(G_{\mu\nu} 45_H G^{\mu\nu}\right)=0$, and the next-to-leading order analysis is possible.

Although this model was abandoned for a long time due to the presence of tachyonic instabilities in the tree-level spectrum, it was found\cite{Bertolini:2009es} that this problem is cured at the quantum level. Subsequently, the SO(10) unification with scalar sector composed of 10-, 45- and 126-dimensional representations was studied in depth and it was observed
that the exact unification and also the seesaw scale $\sigma\equiv\langle 126 \rangle$ in the ballpark of $10^{13}-10^{14}\,\mathrm{GeV}$ may be achieved, if the scalar fields with SM quantum numbers $(8,2,+1/2)$ or $(6,3,1/3)$ are accidentally light. In both cases the scalar spectrum was computed, hence the two-loop analysis including the one-loop matching was performed\cite{Bertolini:2013vta,Kolesova:2014mfa} and an interesting feature of the parameter space compatible with the unification and all the other constraints was found: Either the proton decay will be seen in the Hyper-Kamiokande experiment or only the light octet scenario is viable and this scalar is within the reach of LHC\cite{Kolesova:2014mfa}.

\section{Conclusions}
Some of the obstacles to the precise determination of the proton lifetime in the GUTs were mentioned in this text. Namely, we show that the recent error in the measurement of the strong coupling does not allow the computation beyond the two-loop precision level. Moreover, the Planck-suppressed operators or ignorance of the mass spectrum may introduce an uncertainty comparable with the one-loop error. In this case even the two-loop precision can not be reached and using a toy model we show that the error in the proton lifetime estimate considerably exceeds one order of magnitude, the size of the improvement of the planned experiments with respect to the recent bounds. On the other hand, we presented an SO(10) model where the Planck-suppressed operators are absent because of the group-theory structure, hence the two-loop precision may be achieved.

\section*{Acknowledgements}
The work of H.K. is supported by Grant Agency of the Czech Technical University in Prague, grant No. SGS13/217/OHK4/3T/14.

\section*{References}


\begin{thebibliography}{99}
\bibitem{mine}
S.~Mine.
\newblock Recent results from {S}uper-{K}.
\newblock Talk at Moriond 2015.

\bibitem{Weinberg:1980wa}
Steven Weinberg.
\newblock {Effective Gauge Theories}.
\newblock {\em Phys.Lett.}, B91:51, 1980.

\bibitem{Bertolini:2013vta}
Stefano Bertolini, Luca Di~Luzio, and Michal Malinsky.
\newblock {Light color octet scalars in the minimal SO(10) grand unification}.
\newblock {\em Phys.Rev.}, D87:085020, 2013.

\bibitem{Calmet:2008df}
Xavier Calmet, Stephen~D.H. Hsu, and David Reeb.
\newblock {Grand unification and enhanced quantum gravitational effects}.
\newblock {\em Phys.Rev.Lett.}, 101:171802, 2008.

\bibitem{proton}
P.~Nath and P.F. P\'{e}rez.
\newblock Proton stability in grand unified theories, in strings and in branes.
\newblock {\em Physics Reports}, 441:191--317, 2007.

\bibitem{Aoki:2013yxa}
Y.~Aoki, E.~Shintani, and A.~Soni.
\newblock {Proton decay matrix elements on the lattice}.
\newblock {\em Phys.Rev.}, D89(1):014505, 2014.

\bibitem{Bertolini:2009es}
Stefano Bertolini, Luca Di~Luzio, and Michal Malinsky.
\newblock {On the vacuum of the minimal nonsupersymmetric SO(10) unification}.
\newblock {\em Phys. Rev.}, D81:035015, 2010.

\bibitem{Kolesova:2014mfa}
Helena Kole\v{s}ov\'{a} and Michal Malinsk\'{y}.
\newblock {Proton lifetime in the minimal SO(10) GUT and its implications for
  the LHC}.
\newblock {\em Phys.Rev.}, D90(11):115001, 2014.
%
%
%
%
\end{thebibliography}

\end{document}